\documentclass[12pt]{article}
\pdfoutput=1
\usepackage{amssymb}
\usepackage{graphicx}
\usepackage{graphics}
\usepackage{amsmath}

\def\d{\partial}
\def\l{\left(}
\def\r{\right)}

\newcommand{\be}{\begin{equation}}
\newcommand{\ee}{\end{equation}}
\newcommand{\ba}{\begin{align}}
\newcommand{\ea}{\end{align}}
\newcommand{\bg}{\begin{gather}}
\newcommand{\eg}{\end{gather}}
\newcommand{\bseq}{\begin{subequations}}
\newcommand{\eseq}{\end{subequations}}

\begin{document}

\title{$R^2$-inflation with conformal SM Higgs field}
\author{Dmitry Gorbunov$^{1,2}$, Anna Tokareva$^{1,3}$\\
\mbox{}$^{1}$ {\small\em Institute for Nuclear Research of Russian Academy of
  Sciences, 117312 Moscow,
  Russia}\\  
\mbox{}$^{2}$ {\small\em Moscow Institute of Physics and Technology, 
141700 Dolgoprudny, Russia}\\ 
\mbox{}$^{3}$ {\small\em Faculty of Physics of Moscow State
  University, 119991 Moscow, Russia}
}
\date{}

\maketitle

\begin{abstract} 
We introduce conformal coupling of the Standard Model Higgs field to
gravity and discuss the subsequent modification of $R^2$-inflation. 
The main observation is a lower temperature of reheating which happens mostly
through scalaron decays into gluons due to the conformal (trace) 
anomaly. This modifies all predictions of the original
$R^2$-inflation. To the next-to-leading order in slow roll parameters
we calculate amplitudes and indices of scalar and tensor perturbations
produced at inflation. The results are compared to the next-to-leading
order predictions of $R^2$-inflation with minimally coupled Higgs
field and of Higgs-inflation. We discuss additional features in
gravity wave signal that may help to distinguish the proposed variant
of $R^2$-inflation. Remarkably, the features are expected in the
region available for study at future experiments like BBO and
DECIGO. Finally, we check that (meta)stability of electroweak vacuum in the
cosmological model is consistent with recent results of searches for
the Higgs boson at LHC. 
\end{abstract}

\section{Introduction and Summary}
\label{Sec-I}

The Starobinsky model of inflation \cite{starobinsky} is the first,
yet realistic example of new physics capable of solving  major
problems of the Hot Big Bang theory, see e.g.\,\cite{rubakov2}. It
exploits dynamics of gravity sector, modified by a quadratic in scalar
curvature term added to the gravity action. The attractive feature of
the model is that one and the same force--gravity--is responsible for
both inflation and subsequent reheating of the early Universe. Such 
{\it minimality} is of some interest, given the absence of any direct
evidence of relevant new physics in laboratory and accelerator experiments. 

In this paper we consider the Starobinsky inflation with matter sector
described by the Standard Model of particle physics (SM) which scalar
sector is slightly modified. Namely, we add a conformal coupling of
the SM Higgs field to gravity. This term leaves intact the low energy
phenomenology of the SM, but impacts on the history of the early
Universe. Indeed, we found that with the Higgs boson becoming
conformal at high energies, reheating of the Universe takes place
later and occurs via gluon production due to the conformal (trace)
anomaly. The idea of conformal anomaly being responsible for
reheating was discussed in literature, e.g., 
\cite{Dolgov:1981nw,Watanabe:2010vy}. 
Here it is 
\emph{natural} consequence of the conformal symmetry in our model. 

Lower reheating temperature implies longer matter dominated stage
between inflation and reheating. This modifies all predictions for
power spectra of scalar and tensor perturbations generated at
inflation. Likewise, this modifies predictions for gravity wave signals
expected from nonlinear structure dynamics at post-inflationary stage.
These are special signals in gravity waves given the long-lasting
post-inflationary matter dominated stage.  Remarkably, the signals fall
in the region expected to be reached by proposed future experiments
like BBO\,\cite{BBO} and DECIGO\,\cite{DECIGO} on searches for gravity
waves. These signals have been proposed \cite{bezrukov} as signatures
of $R^2$-inflation. The same is true for our variant with conformal
Higgs, where the features in gravity wave spectrum are expected at different
frequencies, which allows to test the model. Finally, the non-minimal
coupling provides with additional term in the Higgs effective
potential, which becomes important at large scalar curvature. This can
change the answer to the question: in which vacuum does the Higgs
field fall in the expanding Universe, given the value of the Higgs
self-coupling (or the Higgs boson mass)?  

We address all these issues below. The model is presented in
Sec.\,\ref{Sec-II}, and reheating is studied in
Sec.\,\ref{Sec-III}. Predictions for amplitudes and spectral indices
of scalar and tensor perturbations are obtained in
Sec.\,\ref{Sec-IV}. The calculations are performed to the
next-to-leading order in slow roll parameters and the results are
compared to similar predictions obtained there for the original
Starobinsky model and for Higgs-inflation\,\cite{Bezrukov:2007ep}. All
the three models exhibit the same inflationary dynamics, so the only
difference is in reheating temperature, which helps to distinguish the
model predictions. In Sec.\,\ref{Sec-V} we discuss the gravity wave
signals expected in the model: one comes from inflation, others from
nonlinear evolution of inhomogeneities at post-inflationary
matter-dominated stage. Sec.\,\ref{Sec-VI} is devoted to analysis of
stability of electroweak vacuum. There we estimate the lower bound on
the Higgs boson mass, corresponding to the viable cosmological
evolution in the model and find it to be consistent with recent
results of LHC.

\section{The model description}
\label{Sec-II}

The Starobinsky model of inflation is described in the 
Jordan frame by the following action 
\cite{starobinsky,Faulkner:2006ub},
\begin{equation}
\label{original-R2}
S=-\frac{M_P^2}{2}\int \!\! \sqrt{-g}\;d^4x\, \left(
R-\frac{R^2}{6\,\mu^2} \right) + S_{matter}\;. 
\end{equation}
Here the reduced Planck mass is $M_P=M_{Pl}/\sqrt{8\pi}=2.4\times
10^{18}$\,GeV and $S_{matter}$ in our case refers to the SM action. At
large values of scalar curvature $R$ model \eqref{original-R2} allows
for an inflationary stage in a slow roll regime: a scalar degree of
freedom (dubbed {\it scalaron}), emerging when $R^2$-term is added to
the gravity action, plays inflaton.  Scalaron quantum fluctuations
evolving at inflationary stage freeze out with amplitude $\sim\!\mu$
when exit the horizon. Later they give rise to matter perturbations,
which amplitude is fixed by the global fit to cosmological data,
consequently the value of parameter $\mu$ equals \cite{Faulkner:2006ub}
\begin{equation}
\label{mu-value}
\mu=1.3\times 10^{-5}\, M_P=3.1\times10^{13}\,{\rm GeV}\;.
\end{equation}

After inflation the Universe expansion is driven by oscillating
massive scalaron field responsible for the effective
matter-dominated post-inflationary stage. Later, scalaron oscillations
decay into the SM Higgs bosons due to gravity interactions and a hot
stage in the Universe starts with temperature \cite{scalaron}
\begin{equation}
\label{original-temperature}
T_{\rm reh}^{R^2}=3.1\times10^9\,{\rm GeV}\;.
\end{equation}

In this paper we introduce in model \eqref{original-R2} conformal
coupling of the SM Higgs field to gravity. Then the part of action
with the Higgs doublet ${\cal H}$ 
reads (we omit irrelevant for our study Yukawa terms):  
\begin{equation}
\label{conformal-Higgs}
S_H=\int \!\!\sqrt{-g}\; d^4 x \l \frac{1}{6}R\,{\cal H}^{\dagger}{\cal
  H}  + D^{\mu}{\cal H}^\dagger D_{\mu} {\cal H} - \frac{\lambda}{4}\l
{\cal  H}^{\dagger}{\cal H}-v^2\r^2\r \;.
\end{equation}
Note that a non-minimal coupling term is generally 
required by renormalizability of the model in the 
curved space-time, and the particular value of non-minimal coupling is
stable with respect to perturbative quantum corrections. 
After the conformal (Weyl) transformation to the Einstein frame 
\begin{equation}
\label{Weyl-transformation}
g_{\mu\nu}\rightarrow e^{\sqrt{2/3}\,\phi/M_P}g_{\mu\nu} 
\end{equation}
action \eqref{original-R2} takes form \cite{magnano}
\[
S=\int{\!\!\sqrt{-g}\; d^4 x  \l -\frac{M_P^2}{2}R + \frac{1}{2}\d_{\mu} \phi
  \d^{\mu}  \phi - V(\phi)\r} + \tilde{S}_{matter}\;,
\]
\[
V(\phi)=\frac{3\mu^2 M_P^2}{4}  \l 1-e^{-\sqrt{2/3}\phi/M_P}\r^2\;.
\]
Here $\tilde{S}_{matter}$ is the conformally transformed action of
matter (SM fields). Any conformal non-invariance in the matter sector
produces interaction between scalaron $\phi$ and SM particles.

\section{Reheating via the conformal (gauge) anomaly}
\label{Sec-III}

In our model \eqref{original-R2}, \eqref{conformal-Higgs} the SM Higgs
is conformal at large values of the field and hence it decouples from
scalaron. Then the strongest relevant 
coupling between scalaron and SM fields comes 
from the conformal (gauge) anomaly which takes care of reheating in
the model. Indeed, the conformal transformation
\eqref{Weyl-transformation} yields scalaron coupling to the trace of
energy-momentum tensor of matter fields  $T^\mu_\mu$: 
\begin{equation}
\label{coupling-to-trace}
S_{int}=\int{\!\!\sqrt{-g}\; d^4x \, \frac{1}{\sqrt{6}}\,\frac{\phi}{M_P}\, 
T^{\mu}_{\mu}}\;. 
\end{equation}
The relevant terms in \eqref{coupling-to-trace} 
are due to conformal (gauge) anomaly, see e.g. \cite{morozov}: 
\begin{equation}
T^{\mu}_{\mu}=\frac{\beta(\alpha)}{4\,\alpha}\,(F^a_{\mu\nu})^2\,,~~~
\beta\l \alpha\r = \frac{b_\alpha\,\alpha^2}{2\,\pi}\;.
\end{equation} 
Here $F_{\mu\nu}^a$ stand for gauge field tensors and $b_\alpha$ are
the first coefficients of $\beta$-functions for corresponding gauge
coupling constants $\alpha$; for the SM couplings of $U(1)_Y$, $SU(2)_W$ and
$SU(3)_c$ gauge interactions these coefficients are 
$\frac{41}{6}$, $-\frac{19}{6}$ and $-7$, respectively.

As a result, the scalaron decay rate into the gauge bosons 
is 
\begin{equation}
\label{anomaly-decay-rate}
\Gamma_{\phi\to\, {\rm 2\,bosons}} = 
\frac{b_\alpha^2\,\alpha^2 N_{\rm adj}}{768\,\pi^3}\, \frac{\mu^3}{M_P^2}\;,
\end{equation} 
where $N_{\rm adj}=1,\, 3,\, 8$ for $U(1)_Y$, $SU(2)_W$ and
$SU(3)_c$ gauge interactions, correspondingly.  
Values of $\alpha$ must be taken at the scale of $\mu/2$ and we obtain
them by making use of the numerical code \cite{program} operating with
the SM 3-loop $\beta$-functions \cite{Chetyrkin:2012rz}. 

Scalaron decays mostly into gluons, which immediately rescatter
producing all the SM 
particles\footnote{Amplitudes of scalaron direct decays to pair of
  SM 
  massive fermions are suppressed by corresponding
  Yukawa coupling constants and ratio of the Higgs
  field $\sim {\cal H}$, see Sec.\,\ref{Sec-VI}, to the Planck
  scale. Rates of three and four body scalaron decays to SM particles are
strongly  suppressed by coupling constants and the phase space volume.}. 
The scalaron total decay rate
$\Gamma_\phi$ in our model with conformal Higgs field 
\eqref{original-R2}, \eqref{conformal-Higgs} is about 140 times
lower than that in the model with the Higgs field minimally coupled to
gravity \eqref{original-R2}.  
For completeness, let us note that 
if the Higgs non-minimal coupling to
  gravity parameterised as    
$L_{{\rm int}}=\xi\,R\, {\cal H}^{\dagger}{\cal H}$ we obtain for  
the total scalaron decay width:
\be 
\Gamma_{\phi}=\frac{\mu^3}{192\pi M_P^2}\left[\frac{\Sigma b_i^2 \alpha_i^2 N_i^{{\rm adj}}}{4\pi^2}+4(1-6\xi)^2 \right]
\label{f:general}
\ee
where the sum in brackets is taken over the SM gauge groups. 
The decay to gauge fields dominates for $|\xi-1/6|<0.007$. 

We define the reheating temperature of
the Universe after inflation as temperature at the moment of 
equality between the energy densities of  
scalaron condensate and relativistic matter \cite{scalaron}. Then
numerically 
\begin{equation}
\label{conformal-temperature}
T_{\rm reh}=1.1\times g_{*}^{-1/4}\! 
\l T_{\rm reh}\r \sqrt{\Gamma_\phi\, M_P} = 1.4\times
10^8\,{\rm GeV}\;,  
\end{equation}
where effective number of degrees of freedom in the plasma of SM
particles is $g_*\l T_{\rm reh}\r =106.75$. 
For all other values of non-minimal coupling constant $\xi$ in front of the
first term in Eq.\,\eqref{conformal-Higgs} 
but $1/6$ the reheating temperature is higher
than \eqref{conformal-temperature}. In particular, when $\xi$
drops to zero the reheating temperature approaches $3.1\times
10^9$\,GeV \cite{scalaron} as $T_{\rm reh}\propto (1-6\xi)$ provided
Eq.\,\eqref{f:general}. 


\section{Parameters of scalar and tensor perturbations produced at
  the inflationary stage}
\label{Sec-IV}

In both variants of $R^2$-inflation (with ordinary and with conformal
Higgs field) the scalaron potential at inflation is the
same. Therefore, the only difference in predictions for parameters of
scalar and tensor perturbations generated at inflation is due to 
different number of e-foldings because of 
different reheating temperature in the models (for detailed
explanation see e.g. \cite{bezrukov}). Since parameters of perturbation 
power spectra depend on the reheating temperature very
mildly (logarithmically) and the latter differs in the two models
only by factor of 20, cf. Eqs. 
\eqref{original-temperature} and \eqref{conformal-temperature}, they
must be evaluated to the next-to-leading order in slow-roll
parameters. 

The procedure is described e.g. in Refs.\,\cite{Stewart,Liddle}. For
the leading order estimates one usually exploits the number $N_e$ of
e-foldings passed after the perturbation of a given  conformal moment $k$
exited the horizon  \cite{rubakov2,Bezrukov:2008ut}, 
\begin{equation}
\label{e-foldings}
N_e=\log \l\frac{a(k)}{a_e}\r \approx 53.27- 
\frac{1}{3} \log \l\frac{1.4\times 10^{8}\,{\rm
    GeV}}{T_{\rm reh}}\r\;.
\end{equation}
Here $a(k)$ and $a_e$ refer to the scale factor at the moment of
horizon exit and at the end of inflation, respectively. The value of
$k$ is chosen to match the WMAP pivot scale $k/a_0=0.002\,{\rm
  Mpc}^{-1}$, where $a_0$ is the present scale factor. 

For the next-to-leading order calculations a more convenient measure
of the moment of horizon exit is \cite{Liddle} 
\begin{equation}
\label{NLO-e-foldings}
\tilde{N_e}=\log \l \frac{a(k)\,H(k)}{a_e\,H_e}\r\;,
\end{equation}
where $H(k)$ and $H_e$ stand for the Hubble parameter at the exit and
at the end of inflation. The latter is defined as the moment when the
Universe stops to expand with acceleration, i.e. when $d^2 a/dt^2=0$. 
Then one obtains 
\begin{equation}
\label{NLO-number}
\tilde{N_e}=53.80-\frac{1}{3}\log\l 
\frac{1.4\times 10^{8}\,{\rm GeV}}{T_{\rm reh}}\r\;.
\end{equation}
Quantity \eqref{NLO-e-foldings} 
is related to the small slow roll parameters as follows \cite{Liddle} 
\begin{equation}
\label{integral}
\tilde{N_e}\approx - \frac{2\,\sqrt{\pi}}{M_P} \int_{\phi_k}^{\phi_e}
\frac{d\phi}{\sqrt{\epsilon(\phi)}} \l
  1-\frac{1}{3}\,\epsilon(\phi)-\frac{1}{3}\,\eta(\phi)\r\;,
\end{equation}
where $\phi_k$ and $\phi_e$ refer to the moment of horizon exit and
the end of inflation, correspondingly. Introducing variable
$\chi=\exp(\sqrt{2/3}\,\phi/M_P)$ we write down the slow roll
parameters in $R^2$-model (see e.g. \cite{rubakov2}): 
\begin{align}
\label{epsilon}
\epsilon&\equiv \frac{M_P^2}{2}\l\frac{V'}{V}\r^2 
=\frac{4}{3}\frac{1}{(\chi-1)^2}\;,
\\
\label{eta}
\eta&\equiv M_P^2\,\frac{V''}{V} = \frac{4}{3}\frac{2-\chi}{(\chi-1)^2} 
=\epsilon-\frac{2}{\sqrt{3}}\sqrt{\epsilon}\;,
\\
\zeta^2 &\equiv M_P^4\frac{V'V'''}{V^2}=\frac{4}{3}\epsilon\,\l 
1-\frac{3}{2}\sqrt{3\epsilon}\r\;,
\end{align}
where prime denotes derivative with respect to scalaron field
$\phi$. Plugging \eqref{epsilon}, \eqref{eta} into \eqref{integral}
and integrating then Eq.\,\eqref{integral} one extracts slow roll
parameters as functions of 
\[
N\equiv\frac{4}{3}\,\tilde{N_e}+\chi_e-1\;,
\]
namely:
\begin{equation}
\epsilon=\frac{4}{3}\frac{1}{N^2}+O\l \frac{\log(N)}{N^3}\r,
~~\eta=-\frac{4}{3}\frac{1}{N}+\frac{4}{3}\frac{1}{N^2}+O\l
\frac{\log(N)}{N^3}\r, ~~
\zeta=\frac{4}{3}\frac{1}{N}+O\l \frac{\log(N)}{N^3}\r. 
\end{equation}
Numerically, from the Friedman equation 
at the end of inflation, $\chi_e\approx 4.6$, and the
relevant slow roll parameters are reasonably small, $\epsilon_e\approx 0.10$,
$\eta_e\approx 0.27$, which justifies using of approximate relation
\eqref{integral} (see Ref.\,\cite{Liddle} for the exact relation). 

Tilts of the power spectra of scalar and tensor perturbations,
(1-$n_s$) and $n_T$, and tensor-to-scalar ratio $r$ to the
next-to-leading order in the slow roll parameters are given by 
\cite{Stewart,Liddle}
\begin{align}
\label{ns}
1-n_s=6\,\epsilon-2\,\eta-\frac{2}{3}\,\eta^2+0.374\,\zeta^2
&=\frac{8}{3}\frac{1}{N}+\frac{4.813}{N^2}+O\l\frac{\log(N)}{N^3}\r\;,
\\
\label{r}
r=16\,\epsilon&=\frac{64}{3}\frac{1}{N^2}+O\l\frac{\log(N)}{N^3}\r\;,
\\
\label{nT}
n_T=-2\,\epsilon&=-\frac{8}{3}\frac{1}{N^2}+O\l\frac{\log(N)}{N^3}\r\;.
\end{align}
Finally, substituting \eqref{conformal-temperature}, \eqref{NLO-number} into
\eqref{ns}-\eqref{nT} we obtain predictions for the cosmological
parameters. They are presented in Table\,\ref{predictions}, 
\begin{table}[!htb]\centering{
\begin{tabular}{|c|c|c|c|c|}
\hline
Model & $T_{\rm reh}$,\,GeV & $n_s$ & $r$ & $n_T$ \\ 
\hline
$R^2$ with conformal Higgs & $1.4\times 10^8$ & 0.9638 & 0.0038 &
-0.00047 \\ 
\hline
$R^2$ & $3.1\times 10^9$\,\cite{scalaron} & 0.9644 & 0.0036 &
-0.00045 \\ 
\hline
Higgs-inflation & $6\times 10^{13}$\,\cite{Bezrukov:2008ut} & 0.9664 & 0.0032 &
-0.00040 \\
\hline
\end{tabular}}
\caption{Next-to-leading order predictions for spectral parameters. 
\label{predictions}}
\end{table}
together
with predictions for two other inflationary models exhibiting the same
inflaton potential at inflationary stage: $R^2$-model with the Higgs
field minimally coupled to gravity \eqref{original-R2} and the
so-called Higgs-inflation \cite{Bezrukov:2007ep}. For these latter two
models we refine the results of Ref.\,\cite{bezrukov} for $n_s$
derived there to the leading order in slow roll parameters. The
(absolute) error is expected to be of about $1/N^2\simeq 10^{-3}$. 
Indeed, our estimates of
$n_s$ in these models deviate from those in Ref.\,\cite{bezrukov} 
by this amount.
 
For all
the three models the 
interesting values of $n_s$ and $r$ are well inside the region
preferred by combined analysis of present cosmological
data \cite{Beringer:1900zz}. 
Nominally, the (absolute) error of the next-to-leading approximation
$\log(N)/N^3=10^{-5}$ is enough to ensure that it is possible
to distinguish all three models by measuring the fourth digit in the
values $n_s$ and $r$, parameter $n_T$ seems less promising. Provided
the lower reheating temperature, predictions in our model deviate
slightly stronger from those in the Higgs-inflation, as compared to
the predictions in $R^2$-model with the Higgs field minimally coupled
to gravity. Thus, future experiments, like CMBPol
\cite{Baumann:2008aq} with accuracy 
of $10^{-3}$ in $r$ and $0.0016$ in $n_s$, have better chance to
distinguish the Higgs-inflation from our variant of $R^2$, if any
signal would point at the right ballpark.


\section{Gravity wave signals from inflation and from scalaron clumps}
\label{Sec-V}

In this model there are two sources of gravity waves: metric
fluctuations at inflationary stage and scalaron clumps at late
post-inflationary stage. Let us discuss them in order.

\begin{figure}[!htb]
\centerline{
\includegraphics[width=0.8\textwidth]{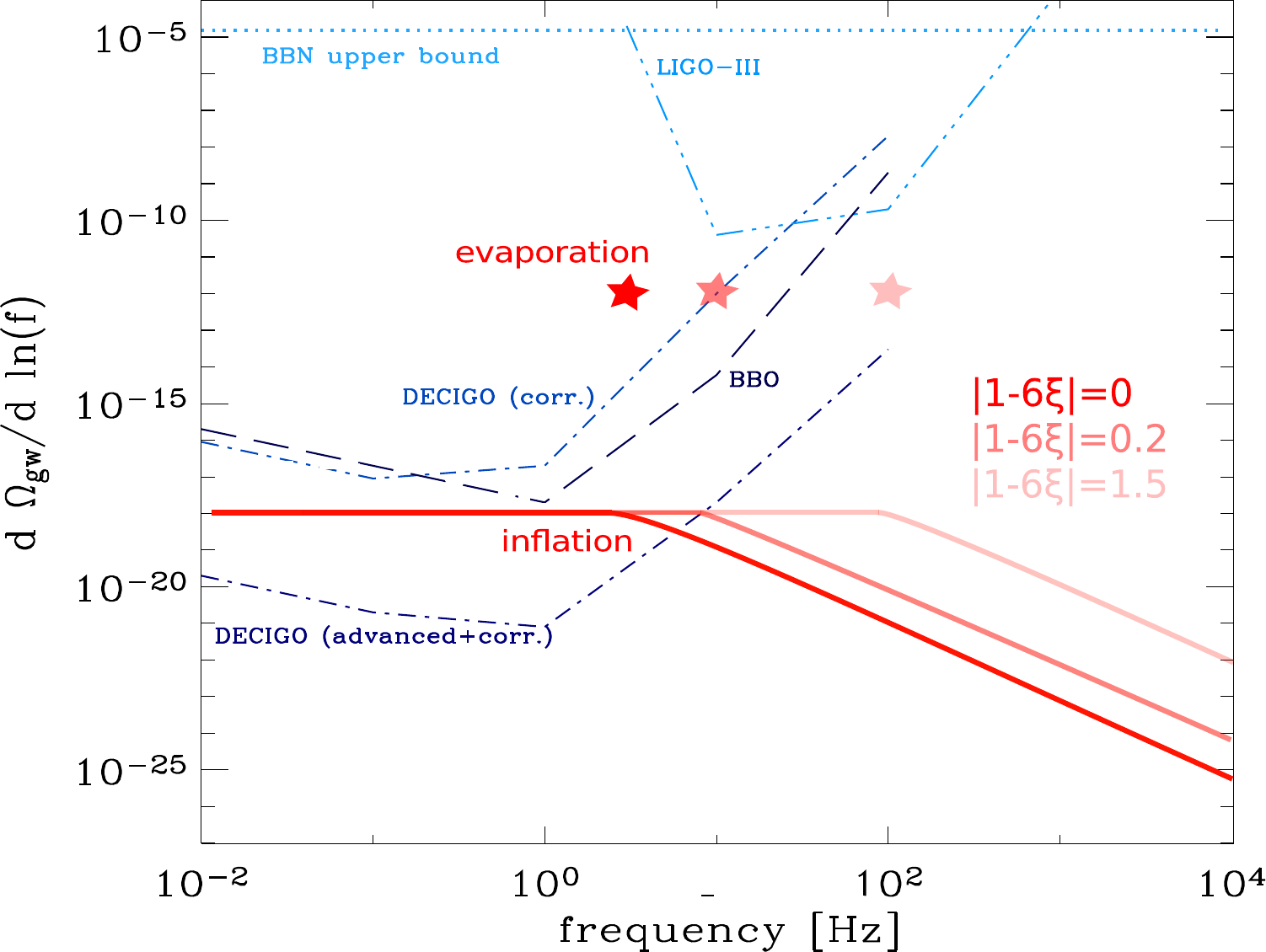}}
\caption{Energy density in gravity waves (in units of the present day
  critical density) $\Omega_{\rm gw}$ 
as a function of frequency and the projected
  sensitivities of next-generation gravitational wave detectors: 
LIGO \cite{LIGO}, BBO \cite{BBO}, 
  DECIGO \cite{DECIGO}. The picture
  shows the gravity wave signal from inflation (solid line) and from
  structure evaporation at reheating (red star); results are presented
  for three particular values of the nonminimal coupling $\xi$. 
\label{GW-plot}
}
\end{figure}

The first source is inherent in any inflationary model.  In the model
under discussion it gives rise to tensor perturbations (gravity waves)
with almost flat power spectrum after inflation: a deviation from
flatness is characterized by index $n_T$ presented in
Table\,\ref{predictions}. The total power in tensor perturbations is
about 0.4\% of that in scalar perturbations, see values of parameter $r$ in
Table\,\ref{predictions}. In the expanding Universe the perturbations,
which length became smaller than horizon, start to evolve. Energy
density of subhorizon tensor modes decreases with scale factor as 
radiation energy density, i.e. as $1/a^4$. Hence at post-inflationary stage when
the Universe is dominated by oscillating scalaron with energy density
scaling as $1/a^3$, the relative contribution of subhorizon gravity waves to
total energy density drops as $1/a$ up to reheating, and later at
radiation domination remains constant. Hence one expects a knee-like
feature in the gravity wave spectrum to be observed at a frequency
$f_*$ determined by the horizon size at reheating $H_{\rm reh}$. The
latter is related to the reheating temperature through the Friedman
equation, so that
\begin{equation}
\label{Hubble-at-reheating}
H_{\rm reh}=\frac{\pi}{\sqrt{90}} 
\frac{g_*\!\l T_{\rm reh}\r\,T^2_{\rm reh}}{M_P}\;.
\end{equation}
The relative contribution of tensor modes of conformal momenta
\[
k>k_{\rm reh}=\frac{H_{\rm reh}}{a_{\rm reh}}
\]
are suppressed. At present $k_{\rm reh}$ corresponds to the physical
momentum 
\begin{equation}
\label{physical-momentum}
p_*=\frac{k_{\rm reh}}{a_0}=\frac{a_{\rm reh}}{a_0}\,H_{\rm reh}\;.
\end{equation}
By making use of entropy conservation we obtain 
\begin{equation}
\label{entropy-conservation}
\frac{a_{\rm reh}}{a_0} = \frac{T_0}{T_{\rm reh}}\l \frac{g_*\l T_0\r}{g_*
\l T_{\rm reh} \r } \r^{1/3}
\end{equation}
with effective degrees of freedom at present 
$g_*\l T_0\r = 3.91$ and $g_*\l T_{\rm reh} \r = 106.75$
\cite{rubakov2}. Then substituting \eqref{entropy-conservation} and 
\eqref{Hubble-at-reheating} into \eqref{physical-momentum} 
we get for the present frequency, where the
knee in gravity wave spectrum is expected, see Fig.\,\ref{GW-plot},  
\begin{equation}
\label{feature}
f_*=\frac{p_*}{2\,\pi}=2.8\, {\rm Hz}\l \frac{T_{\rm reh}}{1.4\times 
  10^8\, {\rm GeV}}\r\;.
\end{equation}

The second source of gravity waves is scalaron inhomogeneities:
subhorizon modes at intermediate matter dominated stage grow 
proportionally to scale factor and have enough time to enter nonlinear
regime before reheating \cite{bezrukov}. Then, gravity waves can be
produced at formation of scalaron clumps, at their subsequent merging and at
final evaporation, see e.g. \cite{1984AmJPh..52..412S,gravity
  waves}. The highest amplitude, hence the best chance to be
observed, is expected from the latter process. Evaporation is the
scalaron decays into relativistic SM particles. It is out-of
equilibrium process yielding a non-vanishing transverse-traceless part
of energy-momentum tensor feeding gravity waves. 
The typical frequency at production is about $H_{\rm reh}$ 
\cite{1984AmJPh..52..412S}, which gets redshifted and presently
coincides with $f_*$ \eqref{feature}.  
The signal amplitude does not depend on the reheating temperature
\cite{1984AmJPh..52..412S}. The estimate in \cite{gravity waves}
gives for a relative contribution of the gravity waves to the present total
energy density $\Omega_{gw}\sim 4 \times 10^{-13}\, \varepsilon$,
where $\varepsilon<1$ is an efficiency factor which represents a
measure of the asphericity of structure evaporation. The possible
signal is shown in Fig.\,\ref{GW-plot}. 

Similar signal is expected\,\cite{bezrukov} 
in $R^2$-model with minimally coupled to gravity Higgs field. However,
the typical frequency is by a factor $T^{R^2}_{\rm reh}/T_{\rm reh}$
higher. The same conclusion holds for position of the knee in gravity
wave spectrum from inflation. One observes in Fig.\,\ref{GW-plot} that
those signals are either at the board of or out of reach of proposed
experiments on searches for gravity waves. The signals in our model
discussed above are right in the region available for investigation by 
future detectors like BBO\,\cite{BBO} and DECIGO\,\cite{DECIGO}, 
see Fig.\,\ref{GW-plot}. This allows for independent
test of our model: signatures in gravity waves pin down the value of
reheating temperature.


\section{Electroweak vacuum of the Higgs potential: lower limits on
  the Higgs boson mass}
\label{Sec-VI}

Since we have modified the Higgs sector by introducing conformal
coupling to gravity, the stability of the electroweak (EW) vacuum  and
whether the Universe ends up in it, given the cosmological evolution
suggested in this paper, have to be investigated. The object under
study is the Higgs effective potential, which in the unitary gauge 
${\cal H}^T=\l 0,\,\l h+v \r /\sqrt{2}\r$ at large $h\gg v=246.2$\,GeV
reads 
\begin{equation}
\label{effective-potential}
V(h)=\frac{\lambda(h)}{4}h^4 - \frac{1}{12}R\,h^2\;. 
\end{equation}
Here $\lambda(h)$ solves the renormalization group equation when $h$
is replaced with renormalization energy
scale\,\cite{Coleman:1973jx,Casas:1994qy}. 
At large $h$ selfcoupling $\lambda(h)$ may
become negative, hence EW vacuum is metastable. 
Coefficient $1/12$ in front of the second
term in \eqref{effective-potential} is renorminvariant (neglecting
graviton loops) \cite{Yoon:1996yr}.  
For homogeneous, isotropic and flat Universe
\begin{equation}
\label{curvature}
R=-12\,H^2-6\,\dot{H}\;,
\end{equation}
where dot denotes the time derivative. Thus, in the hot Universe: at
radiation domination $R=0$, at matter domination $R=-3\,H^2<0$, at
present (dark energy domination), $R<0$. Hence, the conformal coupling
only enlarges the area of stability of the EW (i.e. widens the range of
the allowed Higgs boson mass related to the selfcoupling constant as 
$M_h\approx \sqrt{2\,\lambda}\,v$). 

Actually, metastability of the Higgs potential doesn't mean 
the theory is invalid. The weaker condition to require is the lifetime
of the EW vacuum (with respect to tunneling and thermal decay) 
exceeds the Universe lifetime. For the usual case of minimally coupled 
Higgs the bound coming from tunneling is $M_h\gtrsim 111$\,GeV
\cite{isidori,guidice} which is below the direct lower bound by ATLAS
(115.5\,GeV at 95\% CL\,\cite{ATLAS}). Since in our model the
additional term $R\,{\cal H}^{\dagger}{\cal H}/6$ stabilizes the
potential, the condition from tunneling is fulfilled.

The bound coming from the thermal decay depends on the reheating
temperature \cite{arnold,guidice}. The
numerical calculation (3-loop $\beta$-functions
\cite{Chetyrkin:2012rz} and ${\cal O}\l
\alpha_s^3\r$, ${\cal O}\l\alpha\alpha_s\r$ corrections in matching of
pole and running masses included \cite{program}) 
yields a lower limit which is even below that from tunneling:
\begin{equation}
  M_{therm} = \left[
    109.76
    + \frac{M_t-173.2~{\rm GeV}}{0.9~{\rm GeV}} \times 1.9
    - \frac{\alpha_s-0.1184}{0.0007} \times 0.14
  \right] {\rm GeV}\;.
\end{equation}
Hence, given the direct limit from ATLAS, the problems with EW vacuum
stability in our model are neither at hot stage of the Universe
evolution, nor at present.

In our model the strongest lower limit on the Higgs boson mass comes
from analysis of the Higgs field evolution from inflation to
reheating. There at short periods of time $\Delta t\sim 1/\mu$ 
the scalar curvature is positive because of oscillating scalaron
\eqref{curvature}, 
\begin{equation}
\label{curvature-in-time}
R\simeq - \frac{4}{3\,t^2}\l 1-3\,\cos\l 2\mu t\r\r\;.
\end{equation}  
Then the Higgs potential \eqref{effective-potential} gets negative
mass squared of order $H^2$, which may stimulate escape to large
values of fields and hence to wrong vacuum at late times. Let us
analyze this process. 

Even if initially (classically) at origin, the Higgs field takes 
value $h\sim H$ by the end
of inflation, because of its quantum fluctuations, which get amplified
when became superhorizon. This estimate may be refined, e.g. by making
use of the stochastic approach, see e.g. \cite{fp1,fp2}. 
The comoving
probability $P_c(h,t)$ for the field to take value $h$ at moment $t$
obeys the 
Fokker-Planck equation
\begin{equation}
\label{fp}
\frac{\partial P_c}{\partial t}=\frac{\partial}{\partial h}\left[
\frac{H^3}{8\pi^2}\frac{\partial P_c}{\partial h}
+\frac{V'(h)}{3H}P_c\right]\, .
\end{equation}
One can integrate it over $h$ and introducing the Higgs correlators 
\[
\langle h^2\rangle=\int{\!h^2\,P_c(h,t)\,dh}\;, ~~~~~
 \langle h\,V'(h)\rangle=\int{\!h\,V'(h)\,P_c(h,t)\,dh}
\]
cast it in the following form 
\begin{equation}
\label{chain}
\frac{d}{dt}\langle h^2\rangle =  \frac{H^3}{4\pi^2}
-\frac{2}{3H}\langle h\,V'(h)\rangle\;.    
\end{equation}

Neglecting the $h$-dependence of $\lambda$ and treating perturbations
as Gaussian quantities ($\lambda$ is reasonably small) we simplify
$\langle h\, V^\prime (h)\rangle \simeq 3 \lambda {\langle h^2\rangle}
^2+ 2H^2 \langle h^2\rangle $ and finally arrive at
\begin{equation}
\label{final-fp}
\frac{d}{dt}\langle h^2\rangle=\frac{H^3}{4\pi^2}-\frac{2\lambda }{H}\langle h^2 \rangle^2 - \frac{4 H}{3} \langle h^2 \rangle\;.
\end{equation}
Fluctuation $\langle h^2\rangle$ grows with time due to the first
term and reaches maximum value when the r.h.s. 
of Eq.\,\eqref{final-fp} is zero, that is 
 \begin{equation}
\label{fluctuation}
\sqrt{\langle h^2\rangle_{max}}=\frac{\sqrt{3}}{4\,\pi}\, H\;.
\end{equation}

After inflation the Higgs field starts from\footnote{In fact 
 the resulting lower limit on the Higgs boson mass \eqref{lower-limit}
is (almost) insensitive to the numerical coefficient of order one in
 Eq.\eqref{fluctuation} in front of $H$.} 
\eqref{fluctuation} and evolves according to the classical equation of
motion, 
\begin{equation}
\label{classical-equation}
 \ddot{h}+3H\,\dot{h}+\l -\frac{1}{6}\, R+\lambda(h)\, h^2\r h=0\;,
\end{equation}
where $R(t)$ is given by Eq.\,\eqref{curvature-in-time}. Given
Eq.\,\eqref{curvature-in-time} the last term in
parentheses in \eqref{classical-equation} 
is negligible at small $t$, and the field 
falls from initial value \eqref{fluctuation} as $h\propto t^{-1/3} \propto
1/\sqrt{a}$. However at some moment the potential starts to dominate
($\lambda\, h^2 \propto t^{-2/3}$ falls slower than $R\propto 2/9t^2$) and
if the corresponding value of $h$ is above the maximum of the
effective potential \eqref{effective-potential} then it rolls down
to wrong minimum (towards large $h$). Numerical solution with
$\lambda\l h\r$, evaluated by using 3-loop $\beta$-functions
\cite{Chetyrkin:2012rz} and ${\cal O}\l
\alpha_s^3\r$, ${\cal O}\l\alpha\alpha_s\r$ corrections in matching of
pole and running masses \cite{program}, 
reveals that the Higgs field remains in small
value region (and hence later evolves to the EW vacuum) provided the
Higgs boson mass is above the following critical value
\begin{equation}
\label{lower-limit}
  M_{\rm crit} = \left[
    126.2
    + \frac{M_t-173.2~{\rm GeV}}{0.9~{\rm GeV}} \times 1.55
    - \frac{\alpha_s-0.1184}{0.0007} \times 0.3
  \right] {\rm GeV}\;.
\end{equation}
Uncertainties of estimate \eqref{lower-limit} are associated 
with errors in extraction of SM parameters $M_t$ and $\alpha_s$ from
experimental data (about 2\,GeV at 65\% CL) and unknown higher-order
QCD-corrections to the effective potential
\eqref{effective-potential} (about 1\,GeV) \cite{program}. Given these
numbers, we conclude that obtained limit \eqref{lower-limit} does not
contradict to the recent observation of the SM Higgs-like signals at LHC
\begin{align}
{\rm ATLAS}\;\cite{ATLAS-mass}\;: 
~~~M_h&=126.0\pm0.4(\rm stat.)\pm0.4(\rm syst.)\,GeV\;,\\
{\rm CMS}\;\cite{CMS-mass}\;: 
~~~M_h&=125.3\pm0.4(\rm stat.)\pm0.5(\rm syst.)\,GeV\;.
\end{align}

Hitherto we considered after inflation 
only classical evolution of the Higgs field,  
so a question about its quantum tunneling arises. We have addressed it
adopting the usual instanton approach \cite{Coleman:1977py}. The
initial state for tunneling is a classically evolving Higgs
field. Its rate is of the order of the Universe expansion rate determined by
the Hubble parameter. In the interesting case of our Universe the
tunneling rate is (much) smaller than the expansion rate. 
Thus  we
treat the initial state $h_{in}$ as a stationary state,   
which implies that at $h_{in}$ the Higgs effective 
potential is quite flat, almost reaching the extremum (minimum), 
$V'(h_{in})\approx 0$. 
Then we
approximate the Higgs effective potential by a polynomial of the fourth
degree providing the same positions of the minimum ($h_{in}$,
$V(h_{in})$) and maximum ($h_{max}$, $V(h_{max})$) as the  Higgs
effective potential at 3-loop level we used. 
So the height of the potential barrier remains the
same and it's width becomes 
smaller than for the real potential. We expect
the tunneling rate for the approximate potential to be less than
that in the real case. 

We found the approximate instanton solution sewing 
together polynomial solution inside the new-phase bubble of 
radius $R_b$ with the solution of 
the linear equation outside (neglecting the subdominant in this region
Higgs selfcoupling) and it's Euclidean action $S_E$. 
The tunneling probability per unit time per unit volume is given
by\,\cite{Callan:1977pt}   
\be 
\label{probability}
\Gamma/{\cal V}=D\,\frac{S_E^2}{4\,\pi^2} \,e^{-S_E}\,, 
\ee 
where
dimensionful parameter $D$ comes from the scalar determinant. It is
determined by the size of the tunneling configuration, so we adopt
$D\sim R_b^{-4}$ as an order-of-magnitude estimate.  Then we scanned
over the Higgs mass with step 0.1\,GeV starting from the central value
of Eq.\,\eqref{lower-limit} and using the central values of top mass
and $\alpha_s$. For each Higgs mass value we calculated numerically
the tunneling rate as a function of initial time $t_{in}$ (or Hubble
parameter $H_{in}$) referring to $h_{in}$. The initial state
$h_{in}=h(t_{in})$ was obtained by solving the classical equation of
motion \eqref{classical-equation}.  Requiring for the tunneling rate
in a horizon volume to be always (much) smaller than the Hubble rate
we found the smallest critical mass of the Higgs boson to
be\footnote{The result is almost insensitive to the numerical
  coefficient in the estimate $D\sim R_b^{-4}$ used in
  \eqref{probability}. The tunneling rate changes with the Higgs mass
  mostly because of exponential factor: at critical mass
  \eqref{shifting} the value of $S_E$ jumps by a factor of five.}
\begin{equation}
\label{shifting}
M_{\text{crit}}=126.6\,\text{GeV}\,, 
\end{equation}
that shifts the estimate \eqref{lower-limit} up by $0.4$ GeV. With the
SM Higgs boson mass above the estimate \eqref{shifting} the model is
safe from tunneling to a wrong minima in the early Universe.


We conclude, that being (minimally) extended with free scalar
\cite{Gorbunov:2012ij} or fermion \cite{scalaron} to serve as 
the dark matter, and with sterile neutrinos \cite{scalaron}  to
generate baryon asymmetry of the Universe and explain neutrino
oscillations, the model we discussed becomes a 
phenomenologically complete, yet minimal, viable and testable model of particle
physics.

\hskip 0.5cm

{\bf Acknowledgments. }
The authors are indebted to A.\,Starobinsky and F.\,Bezrukov for
helpful discussions. The work is supported in part by the grant of the
President of the Russian Federation NS-5590.2012.2 and by MSE under
contract \#8412.  The work of D.G. is supported in part by RFBR grant
11-02-01528-a and by SCOPES program.  The work of A.T. is supported in
part by the grants of the President of the Russian Federation
MK-2757.2012.2 and ÌÊ-1754.2013.2.


\end{document}